\documentclass[12pt]{article}
\usepackage{amstex}

\renewcommand{\section}[1] {\vspace{0.6cm}\addtocounter{section}{1}
\setcounter{subsection}{0}\setcounter{subsubsection}{0}
\noindent{\bf\thesection. #1}\par\vspace{0.4cm}}
\renewcommand{\subsection}[1] {\vspace{0.6cm}\addtocounter{subsection}{1}\setcounter{subsubsection}{0}\noindent{\it\thesubsection. #1}\par\vspace{0.4cm}}
\renewenvironment{thebibliography}[1]
{\vspace {0.6cm}\noindent {\normalsize\bf References}
\small\baselineskip=12.8pt\begin{list}
{[\arabic{enumi}]}
{\usecounter{enumi}\setlength{\parsep}{0pt}
\setlength{\leftmargin 1.25cm}{\rightmargin 0pt}
\setlength{\leftmargin 0.75cm}{\rightmargin 0pt}
 \setlength{\itemsep}{3pt} \settowidth
{\labelwidth}{#1.}\sloppy}}{\end{list}}
\input{tcilatex2.5}

\begin{document}

\title{Finite-Difference Implementation of Inviscid Separated Flows with
Infinitely-Long Cusp-Ended Stagnation Zone around Circular Cylinder}
\author{M. D. Todorov \\
%EndAName
{\small \textsl{Dept. of Differential Equations, Institute of Mathematics
and Informatics, }}\\
[-1.mm] {\small \textsl{Technical University of Sofia, Sofia 1756, Bulgaria }%
}\\
[-1.mm] {\small \textsl{e-mail: mtod@@vmei.acad.bg}}}
\date{}
\maketitle

\begin{abstract}
The classical Helmholtz problem is applied for modelling and numerical
investigation of inviscid cusp-ended separated flow around circular
cylinder. Two coordinate systems are used: polar for initial calculations
and parabolic as topologically most suited for infinite stagnation zone.
Scaling by the shape of the unknown free line renders the problem to
computational domain with fixed boundaries. Difference schemes and algorithm
for Laplace equation and for Bernoulli integral are devised. A separated
flow with drag coefficient $C_x=0$ like the so called ``critical'' flow is
obtained. The pressure distribution on the surface of cylinder and the
detachment point compares quantitatively very well with the predictions of
the hodograph method.
\end{abstract}

\section{Introduction}

In 1868 Helmholtz \cite{helmholtz} introduces the notion of a flow
consisting of a potential and stagnant zones matching at priori unknown free
boundaries which are tangential discontinuities and where the balance of
normal stresses (the pressure) holds. Kirchhoff \cite{kirchhoff} came up
with the first solution for the ideal flow around flat plate when the
detachment points were known in advance. Later on in the turning of our
century, Levi-Civita \cite{levicivi}, Villat \cite{villat} etc. developed
further the hodograph method and demonstrated its application to flows
around curved bodies. Satisfying an additional condition for smooth
separation (called now Brillouin-Villat condition \cite
{birkzaran,villat,wu72}) Brodetsky \cite{brode} obtained by the hodograph
method approximate solution for the circular cylinder with a parabolic
expanding at infinity shape of the stagnation zone.

In the years 40 of the present century with the computer advent it was
already possible to calculate such class of flows direct at the physical
plane. The first calculations \cite{imai53,soutvaisy} gave interesting
results. Along with the Brodetsky flow a radically different Helmholtz flow
takes place with decreasing stagnation zone which forms at infinity cusp 
\cite{chaplygin}. Because of  the limitation of computers the shape of the
zone was not conclusive. It appears that the method of hodograph can also be
applied to obtain such a flow (see, e.g., \cite{gurevich}) but only for the
case of circular cylinder. We also found such cusp-ended stagnation zones 
\cite{chritodo84,chritodo86,chritodo87} by means of difference scheme and
confirmed by integral-method calculations \cite{todo93}. A new interesting
solution for the shape of the wake behind the circular cylinder was obtained
after further modifying of the difference scheme. Preliminary results of
this study are represented in \cite{chritodo96}. The features of the
algorithm and our further investigation will be discussed here.

\setcounter{equation}{0}

\section{Posing the Problem}

Consider the steady inviscid flow past a circle -- the cross section of an
infinitely long circular cylinder. The direction of the flow coincides with
the line $\theta = 0,\pi$ of the polar coordinates and the leading
stagnation point of the flow is situated in the point $\theta = \pi$. Taking
into account the symmetry with respect to the line $\theta=0,\pi$ we
consider the flow only in upper half plane.

\subsection{Coordinate Systems}

The gist of our approach is to make use of two different coordinate systems:
the polar one (turning out to be ineffective for the case of infinite
stagnation zones extending far away from the rear end of body) and the
parabolic one the latter being topologically more suited for solving Laplace
equation outside infinitely long stagnation zones. We initiate the
calculations in polar coordinates switching to parabolic coordinates after
the stagnation zone has fairly well developed and has become long enough.

In terms of the two coordinate systems (cylindrical and parabolic) Laplace
equation for the stream function $\psi$ reads: 
\begin{equation}
{\frac{1}{r}}(r\psi_r)_r+ {\frac{1}{r^2}} \psi_{\theta \theta} = 0 \>, \qquad%
\hbox{or}\qquad {\frac{1}{{\sigma^2 + \tau^2 }}} (\psi_{\sigma \sigma} +
\psi_{\tau \tau}) = 0 \>.  \label{eq:govern}
\end{equation}

The undisturbed uniform flow at infinity is given by 
\begin{equation}
\left. \psi \right|_{r \to \infty} \approx r U_\infty \sin{\theta} \>,
\qquad \hbox{or} \qquad \left. \psi\right|_{\sigma \to \infty, \> \tau \to
\infty} \approx \sigma \tau U_\infty \>.  \label{eq:integbc}
\end{equation}

On the combined surface ``body+stagnation zone'' hold two conditions. The
first condition secures that the said boundary is a stream line (say of
number ``zero'') 
\begin{equation}
\psi(R(\theta),\theta) =0, \> \theta \in [0,\pi] \quad \hbox{or} \quad
\psi(S(\tau),\tau) = 0, \> \tau \in (0,\infty) \>,  \label{eq:psizero}
\end{equation}

\noindent where $R(\theta)$, $S(\tau )$ are the shape functions of the total
boundary in polar or parabolic coordinates, respectively. Here and
henceforth we use the notation $\Gamma_1$ for the portion of boundary
representing the cylinder and $\Gamma_2$ -- for the free streamline (Fig.1).

Let $\theta ^{*}$ and $\tau ^{*}$ be the magnitudes of the independent
coordinates for which the detachment of flow occurs. As far as we consider
only the case when the stagnation zone is situated behind the body then the
portion of $\Gamma _{2}$ which describes the free line of the flow is
defined as $0\le \theta \le \theta ^{*}$ or $\tau \ge \tau ^{*},$
respectively. On $\Gamma _{2}$ the shape function $R(\theta )$ is unknown
and it is to be implicitly identified from Bernoulli integral with the
pressure equal to a constant (say, $p_{c}$) which is the second condition
holding on the free boundary. For the two coordinate systems one gets the
following equations for shape functions $R(\theta )$ or $S(\tau )$: 
\begin{eqnarray}
\left[ q+\frac{\psi _{\theta }^{2}}{r^{2}}+\psi _{r}^{2}\right] _{r=R(\theta
)}=1\>, &\qquad \hbox{or}\qquad &\left[ q+\frac{\psi _{\sigma }^{2}+\psi
_{\tau }^{2}}{\sigma ^{^{2}}+\tau ^{^{2}}}\right] _{\sigma =S(\tau )}=1\>.
\label{eq:dyncond} \\
0\le \theta \le \theta ^{*}\>, &\quad &\tau ^{*}<\tau <\infty \>,  \nonumber
\end{eqnarray}
\noindent where $q$ is a dimensionless pressure.

At the symmetry line $\theta=0,\pi$  additional conditions are added

\begin{equation}
\frac{\partial \psi}{\partial \theta} = 0 \>, \ \theta = 0, \pi \qquad%
\hbox{or}\qquad \frac{\partial \psi}{\partial \tau} = 0 \>, \ \tau = 0 \>.
\label{eq:symconds}
\end{equation}

\noindent and thus (\ref{eq:govern}), (\ref{eq:integbc}), (\ref{eq:psizero}%
), (\ref{eq:dyncond}) and (\ref{eq:symconds}) complete b.v.p. for
stream--function $\psi$.

\subsection{Scaled Variables}

The above stated boundary value problem is very inconvenient for numerical
treatment mainly because of two reasons. The first is that the boundary
lines are not coordinate lines. The second is that the shape function of the
stagnation zone must be implicitly identified from the additional boundary
condition (\ref{eq:dyncond}). Following \cite{chrivolk85} we scale the
independent variable ($\theta$ or $\tau)$ by the shape function $R(\theta)$
or $S(\tau)$: 
\[
\eta = rR^{-1} (\theta) \>, \qquad \eta = \sigma -S(\tau).
\]
Such a manipulation renders the original physical domain under consideration
into a region with fixed boundaries, the latter being coordinate lines. In
addition the Bernoulli integral becomes an explicit equation for the shape
function of the free boundary. Scaling the independent variable proved very
efficient in numerical treatment of inviscid or viscous flows with free
boundaries (for details see, e.g., \cite{chritodo86}).

We treat the two coordinate systems in an uniform way denoting $\xi \equiv
\theta$ or $\xi \equiv \tau$ depending on the particular case under
consideration. In terms of the new coordinates $(\eta,\xi)$, the stream
function is a compound function $\tilde \psi (\eta, \theta) \equiv \psi
(r(\eta,\xi),\xi)$ or $\tilde \psi (\eta, \tau) \equiv \psi
(\sigma(\eta,\xi),\xi)$ but in what follows we drop the ``tilde'' without
fear of confusion. The Laplace equation takes then the form 
\begin{equation}
(a \psi_\eta)_ \eta + (b\psi_\xi)_ \xi - (c \psi_\xi)_\eta - (c
\psi_\eta)_\xi = 0  \label{eq:laplace}
\end{equation}

\noindent\raisebox{0.cm}[0.cm][0.cm]{where}\vspace{-0.5cm} 
\[
\begin{array}{c}
a \equiv \eta \bigl[1 + \bigl({\frac{R^{\prime}}{R}} \bigr)^2 \bigr] \>,
\quad b \equiv {\frac{1}{\eta}} \>, \quad c \equiv {\frac{R^{\prime}}{R}} \>;
\\ 
\hbox{or} \\ 
a \equiv 1 + {S^{\prime}}^2, \quad b \equiv 1, \quad c \equiv S^{\prime}\>.
\end{array}
\]
with respective boundary conditions (see \cite{chritodo96}).

Thus we define a well posed boundary value problem for $\psi$ provided that
functions $R(\theta)$ and $S(\tau)$ are known. On the other hand in the
portion $\Gamma_2$ of the boundary (where these functions are unknown) they
can be evaluated from the Bernoulli integral (\ref{eq:dyncond}) which now
becomes an explicit equation for the shape function 
\begin{eqnarray}
{\frac{R^2+{R^{\prime}}^2}{R^4}} \left[\left. {\frac{\partial \bar\psi }{%
\partial \eta}} \right|_{\eta=1} \!\! + R(\theta) \sin \theta \right]^2 = 1
\>, &\ \hbox{or} \ & {\frac{1 + {S^{\prime}}^2}{S^2 + \tau^2}} \left[ \left. 
{\frac{\partial \bar\psi}{\partial \eta}}\right|_{\eta=0} \!\!+ \tau
\right]^2 = 1 \>,  \label{eq:bernoulli} \\
0\le \theta \le \theta^* \>, &\ \phantom{\hbox{or}}\ & \tau^* \le \tau
<\infty \>.  \nonumber
\end{eqnarray}

\setcounter{equation}{0}

\section{Forces Exerted on the Body}

The presence of a stagnation zone breaks the symmetry of the integral for
the normal stresses and hence D'Alembert paradox ceases to exist, i.e. the
force exerted from the flow upon the body is no more equal to zero. Denote
by $\mbox{\boldmath $n$}$ the outward normal vector to the contour $\Gamma$.
Then the force acting upon the contour is given by 
\begin{equation}
\mbox{\boldmath $R$} = - \oint_\Gamma p \mbox{\boldmath $n$} ds = -
\oint_\Gamma (q+p_c) \mbox{\boldmath $n$} ds \ \stackrel{\mathrm{def}}{=} \ {%
\rho aU^2_\infty} \left[C_x\mbox{\boldmath $i$} + C_y\mbox{\boldmath $j$}
\right]\>,  \label{eq:force_integral}
\end{equation}

\noindent where $C_x$ and $C_y$ are the dimensionless drag coefficient and
the lifting force.

After obvious manipulations we obtain for the drag and lifting-force
coefficients the following expression (see \cite{chritodo96}) 
\begin{eqnarray}
C_x = -2\int^{\pi}_{\theta^*} \!\!\! q \left[R(\theta) \cos{\theta} +
R^{\prime}(\theta) \sin{\theta} \right] d\theta &&C_x = 2\int^{\tau^*}_{0}
\!\!\! q \left[ S(\tau) + S^{\prime}(\tau) \tau \right] d \tau  \nonumber \\%
[-0.2cm]
&\hbox{or}  \label{eq:cx_cy} \\[-0.2cm]
C_y &\equiv&0 .  \nonumber
\end{eqnarray}

\noindent where the dimensionless pressure is given by 
\begin{eqnarray}
q = 1 -{\frac{R^2+{R^{\prime}}^2}{R^4}} \left[\left. {\frac{\partial
\bar\psi }{\partial \eta}} \right|_{\eta=1} \!\! + R(\theta) \sin \theta
\right]^2 \!\! &\ \hbox{or}\ & q = 1 - {\frac{1 + {S^{\prime}}^2}{S^2 +
\tau^2}} \left[ \left. {\frac{\partial \bar\psi}{\partial \eta}}%
\right|_{\eta=0} \!\!+ \tau \right]^2 \!\!\!\!\>.   \label{eq:pressure}
\end{eqnarray}

\setcounter{equation}{0}

\section{Difference Scheme and Algorithm}

\noindent\vspace{-0.6cm}\noindent

\subsection{Splitting scheme for Laplace equation}

\nobreak\noindent For the purposes of the numerical solution, the
transformed domain must be reduced to finite one after appropriately
choosing the ``actual infinities''. In the case of polar coordinates the
domain is infinite with respect to coordinate $\eta$ only and it fully
enough to select sufficiently large number $\eta_\infty$ and to consider the
rectangle: $[0\!\le\!\theta\!\le\! \pi; 1\!\le\!\eta\!\le\!\eta_\infty]$
(Fig.2a). In the case of parabolic coordinates an actual infinity is to be
specified also for the $\tau$-coordinate, namely $\tau_\infty$ and to
consider the rectangle: $[0\!\le \!\tau\! \le\!\tau_\infty; \>
0\!\le\!\eta\!\le\! \eta_\infty ]$ (Fig.2b). In both directions we employ
non-uniform mesh. The first and the last $\eta$-lines are displaced
(staggered) from the respective domain boundaries on a half of the adjacent
value of the spacing. Thus on two-point stencils second-order approximation
for the boundary conditions is achieved (see \cite{chritodo86}. The
non-uniformity of the mesh enables us to improve the accuracy near the body
and to reduce the number of points at infinity.

In $\theta$-direction the mesh is not staggered but it is once again
non-uniform being very dense in the vicinity of the rear stagnation point,
i.e. in the vicinity of $\theta = 0$ which is of crucial importance when
acknowledging the infinity in cylindrical coordinates. It is desirable to
have the ``actual infinity" in cylindrical coordinates as larger as possible
in order to prepare the ground for switching to the parabolic coordinates.
The connection between the $\tau$-mesh and $\theta$-mesh is derived on the
basis of the connections between the two coordinate systems, namely 
\begin{equation}
\tau_j = \sqrt{R(\theta_j)\cos{\theta_j} + R(\theta_j)}\>, \quad \hbox{if}%
\quad 0\le \theta_j \le \pi \>; \qquad S_j =\sqrt{2R(\theta_j) - \tau_j^2}
\>,  \label{eq:polar_to_prb}
\end{equation}

\noindent and these relations can be transformed when necessary to calculate 
$S_j, \tau_j$ from $R_j, \theta_j$ or vice versa.

Due to the topological differences between the polar and parabolic
coordinate systems after the transition to parabolic coordinates it is
necessary to generate a new $\tau$-mesh. The new mesh has to be sparse at
large distances behind the body where the gradients of the flow are small.
To this end the knots $\tau_j$ are obtained from (\ref{eq:polar_to_prb})
making use of spline interpolation. The new $\tau$-mesh is uniform on the
rigid body and is changing behind the body according to the quadratic rule 
\begin{equation}
\begin{cases} \tau_j = (j-1)h \>, & h = \frac{3\sqrt{2}}{N}, \ j =
1,\ldots,[\frac{N}{3}]+1, \\ \tau_j = \exp {(j-[\frac{N}{3}]-1)h \ln
\tau_\infty} \>, & h = \frac{3}{2N}, \  j = [\frac{N}{3}]+2, \ldots,N+1
\end{cases}  \label{eq:tau_mesh}
\end{equation}

\noindent where $[\frac{N}{3}]$ is the last point of the rigid body

We solve the boundary value problem iteratively by means of the method of
splitting of operator. Upon introducing fictitious time we render the
equation to parabolic type and then employ the so-called scheme of
stabilising correction \cite{yanenko}. On the first half-time step we have
the following differential equations ($\Delta t$ is the time increment) 
\begin{gather}
\frac{\psi^{n+\frac{1}{2}}_{ij}-\psi^n_{ij}}{\frac{1}{2}\Delta t}={%
\Lambda_2(b\Lambda_2 \psi^{n+\frac{1}{2}})}_{ij} + {\Lambda_1(a\Lambda_1%
\psi^n)}_{ij} - {\Lambda_1(c\Lambda_2 \psi^n)}_{ij} - {\Lambda_2(c\Lambda_1%
\psi^n)}_{ij}  \label{eq:frst_step}
\end{gather}
for $i=2,\cdots,M, \ j=2,\ldots,N$ with respective boundary conditions \cite
{chritodo86}

The second half-time step consists in solving the following differential
equations 
\begin{gather}
\frac{\psi^{n+1}_{ij}-\psi^{n+\frac{1}{2}}_{ij}}{\frac{1}{2}\Delta t}={%
\Lambda_1(a\Lambda_1 \psi^{n+1})}_{ij} - {\Lambda_1(a\Lambda_1\psi^n)}_{ij}
\label{eq:sec_step}
\end{gather}

\noindent for $i=2,\ldots,M, \ j=2,\ldots,N$ with respective boundary
conditions \cite{chritodo86}.

Thus the b.v.p. for the stream function is reduced to consequative systems
with sparse (tridiagonal) matrices (for detail see e.g.,\cite{chritodo86}.
The main advantage of the economical schemes of the splitting type is that
on each half-time step we solve one-dimensional problems with sparse
(tridiagonal) matrices. This can be done by means of the Thomas algorithm 
\cite{roache}.  However, the system for streamfunction $\psi(\eta,\xi)$
cannot be solved by plane Thomas algorithm since the condition for numerical
stability of the elimination is not satisfied for all points of domain. For
this reason a modification of the Thomas algorithm (in fact Gaussian
elimination with pivoting for three-diagonal systems) called
``non-monotonous progonka'' (see \cite{samnik}, \cite{chreding}) is employed
for its solution.

To calculate afterwards the forces acting upon the body we use the simple
formulas for numerical integration based on the trapezoidal rule, which are
consistent with the overall second-order approximation of the scheme.

\subsection{Difference Approximation for the Free Boundary}

The equations (\ref{eq:bernoulli}) can be resolved for the derivatives $%
R^{\prime}(\theta)$ or $S^{\prime}(\tau)$ when the radicals exist, i.e.
following conditions are satisfied: 
\begin{eqnarray}
&&Q(\theta) \stackrel{\mathrm{def}}{=} {\frac{R^2(\theta ) }{T^2(\theta)}} >
1 \>, \ T(\theta) = \left. \frac{\partial \bar\psi}{\partial \eta}
\right|_{\eta=1} \!\! + R(\theta) \sin\theta  \nonumber \\[-0.2cm]
\hbox{or}&&  \label{eq:posit_cond} \\[-0.2cm]
&& Q(\tau) \stackrel{\mathrm{def}}{=} {\frac{S^2(\tau) + \tau^2 }{T^2(\tau) }%
} > 1 \>, \ T(\tau) = \left. \frac{\partial \bar\psi}{\partial \eta}
\right|_{\eta=0} \!\! + \tau\>,  \nonumber
\end{eqnarray}

where connection between functions $T$ is determined simply by the formula

\begin{equation}
T(\theta)={\frac{1}{2}}(S(\tau)-S^\prime \tau)T(\tau)  \label{eq:conTT}
\end{equation}

\noindent The above inequalities are trivially satisfied in the vicinity of
the leading-end stagnation point inasmuch as that for $\theta\rightarrow\pi$
(or $\tau \rightarrow 0$) one has $T \to 0$ and hence ${\frac{R^2}{T^2}}%
\to\infty$ or ${\frac{S^2 + \tau^2 }{T^2 }} \to \infty$. In the present work
we use the dynamic condition (\ref{eq:dyncond}) in polar coordinates only,
so that we present here just the relevant scheme in polar coordinates
without going into the details for parabolic coordinates.

Suppose that the set functions $\psi^\alpha_{ij}, R^\alpha_j, S^\alpha_j,
T^\alpha_j$ are known from the previous global iteration, say of number $%
\alpha $.\footnote{%
We distinguish here between global and local iteration, the latter referring
to the time-stepping of the coordinate splitting method.} We check the
satisfaction of (\ref{eq:posit_cond}) beginning from the point $\theta=0$
and continue with increasing $\theta$ . Let $j^*+1$ be the last point where (%
\ref{eq:posit_cond}) is satisfied and, respectively $j^*$ -- the first one
where it is not (polar coordinates).  The position $\theta^*$ of the
detachment point is captured by means of a linear interpolation 
\[
\theta^* = {\frac{\theta_{j^*+1}q_{j^*} - \theta_{j^*}q_{j^*+1} }{q_{j^*} -
q_{j^*+1}}} \quad \rightarrow \quad g^* = \theta^* - \theta_{j^*+1} \>. 
\]

For the shape function $\hat R_j$ of free line is solved the following
difference scheme 
\begin{gather}
\hat R_{j-1} - \hat R_j = g_j {\frac{\hat R_j+\hat R_{j-1}}{2}} \sqrt{\frac{1%
}{2}\left[\left({\frac{R^\alpha_j }{T^\alpha_j}}\right)^2 + \left({\frac{%
R^\alpha_{j-1} }{T^\alpha_{j-1}}}\right)^2 \right] - 1}  \label{eq:fs_polar}
\end{gather}

\noindent for $j=j^*, \ldots ,2$ , whose approximation is $O(g^2_j)$. Only
in the detachment point the difference scheme is different, specifying in
fact the initial (``inlet'') condition, namely 
\begin{eqnarray}
\hat R_{j^*} - R(\theta^*) = g^* {\frac{R(\theta^*)+ \hat R_{j^*}}{2}} \sqrt{%
\frac{1}{2}\left[\left({\frac{R^\alpha_{j^*} }{T^\alpha_{j^*}}}\right)^2
+\left({\frac{R(\theta^{*})}{T(\theta^*)}}\right)^2\right] - 1}\>,
\end{eqnarray}

\noindent where $R$ without a superscript or ``hat'' stands for the known
boundary of rigid body. Thus the mere condition for existence of the square
root of the Bernoulli integral defines at each iteration stage $\alpha$ the
new approximation for the position of the detachment point so that it
'slides' during the iterations alongside the rigid body. This manner of
determining of the detachment point we called Christov's algorithm (see \cite
{todo93}).

In the end a relaxation is used for the shape-function of the free boundary
at each global iteration according to the formula: 
\[
R^{\alpha+1} = \omega\hat R_j + (1-\omega)R^\alpha_j
\]

\noindent where $\omega$ is called relaxation parameter.

\subsection{The general Consequence of the Algorithm}

\nobreak\noindent Each global iteration contains two stages. On the first
stage, the difference problem for Laplace equation is solved iteratively
either in polar or in parabolic coordinates (depending on the development of
the stagnation zone). The internal iterations (time steps with respect to
the fictitious time in the splitting procedure) are conducted until
convergence is achieved in the sense that the uniform norm is lesser than $%
\varepsilon_2 = 10^{-6}$. Thus the new iteration for stream function $%
\psi^{\alpha+1}_{ij}$ is obtained.

The polar coordinates appear to be instrumental only on the first several
(7-10) global iterations. When the rearmost cusp point of the stagnation
zone reaches 30--50 diameters of cylinder (calibers), the current-iteration
values of the sought functions are transformed to parabolic coordinates and
hence the calculations for the stream function continue solely in terms of
parabolic coordinates.

The second stage of a global iteration consists in solving the difference
problem for the free surface in polar coordinates. The transition to and
from parabolic coordinates is done according to (\ref{eq:polar_to_prb}) and (%
\ref{eq:conTT}). Note that there is one-to-one correspondence between the
points in polar and parabolic coordinates and hence between the respective
values of the scalar set functions $\psi$ and $R$.

The criterion for convergence of the global iterations is defined by the
convergence of the shape function as being the most sensitive part of the
algorithm, namely the global iterations are terminated when 
\begin{equation}
\max_j\big| {\frac{R_j^{\alpha+1}-R_j^\alpha }{R_j^{\alpha+1}}} \big|<
10^{-4}.  \label{eq:normR}
\end{equation}

The obtained solutions for the stream function and the shape function of the
boundary are the values of the last iteration $\psi_{ij} =
\psi^{\alpha+1}_{ij}$ and $R_j = R^{\alpha+1}_{j}$, respectively. Then the
velocity, pressure, and the forces exerted from the flow upon the body are
calculated.

\setcounter{equation}{0}

\section{Results and Discussion}

The numerical correctness of scheme (\ref{eq:frst_step}), (\ref{eq:sec_step}%
) is verified through exhaustive numerical experiments and through
comparison with the known exact solution for the inviscid non-separated flow
past a circular cylinder 
\begin{equation}
\psi = U_\infty (r - \frac{1}{r})\sin \theta \>,  \label{eq:an_sol}
\end{equation}

\noindent where $\psi $ is the stream function, $U_{\infty }$ -- the
velocity of the main flow and $r$ and $\theta $ - the polar coordinates of a
point of the flow. We used different meshes with sizes $M\times N$ : $%
41\times 68$, $41\times 136$, $161\times 158$, $101\times 201$, $101\times
136$, etc. Respectively, the actual infinity $\eta _{\infty }$ assumed in
the numerical experiments the values 5, 10, 20. The dependence of the
numerical solution on the time increment $\Delta t$ is also investigated and
it is shown that the approximation of the stationary part of the equations (%
\ref{eq:frst_step}) and (\ref{eq:sec_step}) does not depend on $\Delta t$,
i.e. the scheme has the property called by Yanenko \cite{yanenko} \textit{%
full approximation}. The relative differences for $\psi $ when $\Delta t$ is
in the interval [0.001,2] do not exceed 0.5\%. The numerical experiments
show that the optimal values for $\Delta t$ is in interval [0.5,1]. For this
reason the rest of the calculations in the present work are performed with $%
\Delta t=0.5$. The comparison of the solution (\ref{eq:an_sol}) to the
present numerical results is quantitatively very good. The deviations for
the different meshes are in order of approximation $O(h^{2}+g^{2})$ and do
not exceed 3\%. For example in case of mesh $161\times 156$ the relative
error is about 0.9\%.

The adequate choice of the ``actual infinities'' $\eta_\infty, \tau_\infty$
and the spacings $h_i, g_j$ have a profound impact on the accuracy of the
difference schemes (\ref{eq:frst_step}) and (\ref{eq:sec_step}). For a given
``actual infinity'' the improvement in the accuracy can be achieved through
increasing the number of mesh points (decreasing the size of spacing). This
makes the use of uniform mesh ineffective because in the far-field region
the gradients of the flow are small and the high resolution is not
necessary. That was the reason to employ the non-uniform meshes. The
``optimal'' value for the relaxation parameter turned out to be $\omega=0.01$%
. Smaller values increased intolerable the computational time while $\omega
> 0.1$ could not ensure the convergence of the global iterations.
Respectively $\eta_\infty=10$ is the optimal value for the lateral ``actual
infinity''

In order to compare calculated results with the prescription of the
Levi-Civita method in case of so called 'critical' separation angle $%
\theta_{*} = 124.2^o$ (in respect to leading stagnation point of the
cylinder) it is necessary to summarize that method and deduce corresponding
relations. Following \cite{birkzaran} the physical plane $z$ is mapped on
the unit halfcircle $t$ so, that free boundary transforms into the diameter
and rigid boundary - into the halfcircumference $t=e^{i \sigma},
0\le\sigma\le\pi$. Then 
\begin{equation}
z={\frac{M }{4}} \int_i^t {e^{i \Omega(t)}(1-it)^2\left(1-{\frac{1 }{t^2}}%
\right){\frac{dt }{t}}}\>,  \label{eq:levichi}
\end{equation}

\noindent where the function $\Omega(t)=\Theta(t)+i\mathrm{T}%
(t)=\sum_{k=0}^\infty a_{2k+1}t^{2k+1}$. Hence we obtained the following
parametrical equations describing boundary of the cylinder from the leading
stagnation point to the separation point $\theta_*$:

\begin{eqnarray}  \label{eq:rigbody}
x_{cyl}(s)&\!\!\!=\!\!\!&\mathbf{Re} z= -M \int_{\frac{\pi }{2}}^s e^{-%
\mathrm{T}(t)}\sin\Theta (t) \sin \sigma (1+\sin\sigma)d\sigma  \nonumber \\%
[-0.2cm]
\\[-0.2cm]
y_{cyl}(s)&\!\!\!=\!\!\!&\mathbf{Im} z= M \int_{\frac{\pi }{2}}^s e^{-%
\mathrm{T}(t)}\cos\Theta (t) \sin\sigma(1+\sin\sigma) d\sigma\>,  \nonumber
\end{eqnarray}

\noindent where ${\frac{\pi }{2}} \le s \le \pi\>,\>
\Theta(t)=\sum_{k=0}^{12} a_{2k+1}\cos(2k+1)\sigma\>,\>\mathrm{T}%
(t)=\sum_{k=0}^{12} a_{2k+1}\sin(2k+1)\sigma\>$, and parametrical equations
describing the freestreamline from the separation point $\theta_*$ to
infinity:

\begin{eqnarray}
x(s)&\!\!\!=\!\!\!&\mathbf{Re} z= {\frac{M }{4}} \int_{-1}^s \left[\left({%
\frac{2 }{t}}- t - {\frac{1 }{t^3}}\right)\cos\Theta(t) -\left({\frac{2 }{t^2%
}} -2\right) \sin\Theta (t) \right] dt + x_{cyl}(\pi)  \nonumber \\
[-0.2cm]  \label{eq:freestr} \\[-0.2cm]
y(s)&\!\!\!=\!\!\!&\mathbf{Im} z = {\frac{M }{4}} \int_{-1}^s \left[\left({%
\frac{2 }{t}}- t - {\frac{1 }{t^3}}\right)\sin\Theta (t) +\left({\frac{2 }{%
t^2}} -2\right) \cos\Theta (t)\right] dt + y_{cyl}(\pi)\>,  \nonumber
\end{eqnarray}

\noindent where $-1 \le s \le 0\>,\>\Theta(t)= \sum_{k=0}^{12} a_{2k+1}
t^{2k+1}$. If the parameters have values $M=5.71464\>,\> a_1=2\>,\>
a_3=.12518\>,\> a_5=.02661\>,\> a_7=.00858\>,\> a_9=.00349\>,\>
a_{11}=.00167\>,\> a_{13}=.00089\>,\> a_{15}=.00053\>,\> a_{17}=.00035\>,\>
a_{19}=.00024\>,\> a_{21}=.00018\>,\> a_{23}=.00016$, it corresponds to the
so called critical separated flow, which detaches at angle $\theta_*$. This
Helmholtz flow has decreasing (concave) stagnation zone with cusp end at
infinity (Chaplygin--Kolscher flow).

Further the velocity

\begin{equation}
v(z)={\frac{1+it }{1-it}}e^{-i\Omega(t)}\>,\quad \hbox{from where}\quad
|v(z)|={\frac{cos \sigma }{1+sin \sigma}} e^{\mathrm{T}(t)}\>,
\label{eq:levchipres}
\end{equation}

\noindent whence it follows immediately that the pressure on the cylinder is 
\begin{equation}
p(\theta)=1-|v|^2\>,  \label{eq:preslevch}
\end{equation}
\noindent where $\theta=\arctan {\frac{y_{cyl}(s) }{1-x_{cyl}(s)}}$ is the
polar angle. %\setcounter{figure}{2}

In Figs.3-a,b are presented the obtained shapes of the stagnation zone
behind the cylinder and in the near wake for resolutions $41\times 68$, $%
81\times 136$ and $101\times 201$ and different values of relaxation
parameter: $\omega = 0.01; 0.001$. Obviously there is an excellent
comparison between different numerical realizations. On the same figure is
added the shape of the Chaplygin--Kolscher flow. The latter we calculate by
means of parametrical equations (\ref{eq:rigbody}), (\ref{eq:freestr}) using
the usual trapezoidal rule. %\cite{birkzaran,gurevich}.
The symbols stand for the results taken from the charts of the paper \cite
{soutvaisy}. It is worth noting the perfect coincidence of the computed by
us separation angle with both this one, computed in \cite{soutvaisy} and the
'critical' one, prescribed by the hodograph (Levi--Civita's) method.
Nevertheless the difference between our solution and this in \cite{soutvaisy}
is sizable due to the inconclusive character of the latter. The logarithmic
scale is used in Fig.3-b in order to expand the differences between the
different difference solutions making them visible in the graph. The shapes
of the free boundary obtained on the three grids with different resolution
are compared among themselves very well up to 200 calibers . It is clearly
seen up to 70 calibers the shapes are practically indistinguishable and up
to 160--170 calibers the relative difference does not exceed 1--3\%
respectively. This supports the claim that indeed a solution to the
Helmholtz problem has been found numerically by means of the developed in
the present work difference scheme. At the Fig.3-b it is seen the
quantitative difference between our numerical solution of cusp-ended type
and this one prescribed by the hodograph method. Indeed there is excellent
agreement concerning the positions of detachment point and pressure
distribution but the hodograph method postulates the asymptotic behaviour of
the free line also. On the contrary we do not set any condition at infinity.
In a sense our free boundary has an implicit numerical ``closure'' of
cusp-ended type.

The calculated dimensionless pressure $q$ is shown in Fig.4. Here is seen
again an excellent comparison among the different mesh resolutions. In the
stagnation zone it is in order of $10^{-4}$, which is in very good agreement
with the assumption that the unknown boundary is defined by the condition $%
q=0$. The amplitude of the minimum of $q$ is smaller than 3 the latter being
the value for ideal flow without separation. This means that the stagnation
zone influences on the flow upstream. On the same figure is presented the
pressure, calculated by means of (\ref{eq:preslevch}) which corresponds to
the separation angle $\theta_{*} = 124.2^o$. Apparently obtained here
pressure approximates very well this curve. It is known the
Chaplygin-Kolscher flow has a vanishing drag coefficient \cite
{birkzaran,gurevich}). In other words there exists an inviscid separated
flow submitted to The D'Alembert paradox like nonseparated. Varying the mesh
parameters we obtained for the drag coefficient $C_x$ values between $%
2\times10^{-2}$ when resolution is $41\times 68$ and $5\times10^{-4}$ when
resolution is $101\times 201$. That is to say our $C_x \approx 0$ and the
error is in order of approximation. In order to confirm the above assumption
we made the following numerical experiment: in formula (\ref{eq:cx_cy}) for
the drag coefficient we replaced our pressure by the pressure obtained from (%
\ref{eq:preslevch}). The calculated value is $C_x = 3\times10^{-4}$.

\section{Concluding Remarks}

An algorithm for numerical solving the classical Helmholtz problem behind a
circular cylinder is developed. Scaled coordinates are employed rendering
the computational domain into a region with fixed boundaries and
transforming the Bernoulli integral into an explicit equation for the shape
function. The crucial feature of the method developed here is that the
detachment point is not prescribed in advance. Rather it is defined
iteratively. Difference scheme using coordinate splitting is devised.
Exhaustive set of numerical experiments is run and the optimal values of
scheme parameters are defined. Results are verified on grids with different
resolutions. The drag coefficient of the calculated separated flow vanishes
like cusp-ended infinite flow obtained by means of the hodograph method.

\bigskip \noindent \textbf{Acknowledgments\/} The author presents his
gratitudes to Prof. C.I. Christov for the many helpful discussions and
support allowing this work to be carried out.

The financial support by the National Science Foundation of Bulgaria, under
Grant MM-602/96 is gratefully acknowledged.

\bigskip \bigskip \centerline{\bf FIGURE CAPTIONS}

\bigskip \centerline{\it fig1.gif}

\centerline{Figure 1: Posing the problem}

\bigskip \centerline{\it fig2a.gif}

\centerline{Figure 2a}

\bigskip \centerline{\it fig2b.gif}

\centerline{Figure 2b}

\bigskip \centerline{\it cylsnear.gif}

\centerline{(a) behind the cylinder}

\bigskip \centerline{\it cylsfar.gif}

\centerline{(b) far wake}

\noindent Figure 3: The obtained separation lines for relaxation parameter $%
\omega = 0.01$ and different resolutions: - - - - $41\times 68$; --- --- --- 
$81\times 136$; -- -- -- $101\times 201$; ------ hodograph method; $%
\triangleright\ \triangleright\ \triangleright\ \triangleright\ $ \cite
{soutvaisy}

\bigskip \centerline{\it prescyls.gif}

\noindent Figure 4: The pressure distribution for relaxation parameter $%
\omega = 0.01$ and different resolutions: - - - - $41\times 68$; --- --- --- 
$81\times 136$; -- -- -- $101\times 201$; ------ hodograph method; $\circ\
\circ\ \circ\ \circ\ $ nonseparated inviscid flow.

\end{document}